\documentclass[11pt,a4paper,twoside]{article}
\usepackage[table]{xcolor}
\PassOptionsToPackage{table}{xcolor}
\usepackage[utf8]{inputenc}
\usepackage{amsmath,amssymb,amsfonts,mathtools}
\usepackage{amsmath}
\usepackage{graphicx}
\usepackage{float}
\usepackage[colorlinks=true,allcolors=blue]{hyperref}
\usepackage[margin=2.5cm]{geometry}
\usepackage{booktabs}
\usepackage{tikz}
\usepackage{pgfplots}
\pgfplotsset{compat=newest}
\usetikzlibrary{arrows.meta,positioning,backgrounds}
\usepackage{subcaption}
\usepackage{siunitx}
\usepackage{color}
\usepackage{fancyhdr}
\usepackage{microtype}
\usepackage{titlesec}
\usepackage[most]{tcolorbox}
\usepackage{enumitem}
\usepackage{multicol}
\usepackage{array}
\usepackage{mathrsfs}
\usepackage{natbib}
\usepackage{tcolorbox}
\usepackage{listings}
\usepackage{algorithm}
\usepackage{algorithmic}
\titleformat{\section}{\normalfont\Large\bfseries}{\thesection}{1em}{}
\titleformat{\subsection}{\normalfont\large\bfseries}{\thesubsection}{1em}{}
\titleformat{\subsubsection}{\normalfont\normalsize\bfseries}{\thesubsubsection}{1em}{}

\definecolor{lightgray}{gray}{0.95}
\definecolor{lightblue}{rgb}{0.93,0.95,1.0}
\definecolor{darkblue}{rgb}{0.1,0.1,0.7}
\tcbset{
enhanced,
colback=lightblue,
colframe=darkblue,
arc=2mm,
boxrule=0.5pt,
fonttitle=\bfseries
}
\title{\Large\textbf{Analysis of COVID-19 Infection Dynamics: \ Extended SIR Model Approach}}
\author{%
\textsc{Caleb Traxler}$^{1}$ \and
\textsc{Minh Ton}$^{1}$ \and
\textsc{Nameer Ahmed}$^{1}$ \and
\textsc{Sasha Prostota}$^{1}$ \and
\textsc{Annie Cheng}$^{1}$ \
\vspace{0.5em}
\small $^{1}$Department of Mathematics, University of California, Los Angeles
}
\date{\today}
\begin{document}
\maketitle
\begin{abstract}
\noindent
This paper presents a detailed mathematical investigation into the dynamics of COVID-19 infections through extended Susceptible-Infected-Recovered (SIR) and Susceptible-Exposed-Infected-Recovered (SEIR) epidemiological models. By incorporating demographic factors such as birth and death rates, we enhance the classical Kermack–McKendrick framework to realistically represent long-term disease progression. Using empirical data from four COVID-19 epidemic waves in Orange County, California, between January 2020 and March 2022, we estimate key parameters and perform stability and bifurcation analyses. Our results consistently indicate endemic states characterized by stable spiral equilibria due to reproduction numbers ($R_0$) exceeding unity across all waves. Additionally, the inclusion of vaccination demonstrates the potential to reduce the effective reproduction number below one, shifting the system towards a stable disease-free equilibrium. Our analysis underscores the critical role of latency periods in shaping epidemic dynamics and highlights actionable insights for public health interventions aimed at COVID-19 control and eventual eradication.
\end{abstract}
\section{Introduction}
\label{sec:introduction}
The COVID-19 pandemic has presented unprecedented challenges to global public health, highlighting the importance of mathematical modeling for understanding disease dynamics and informing intervention strategies. The Susceptible-Infected-Recovered (SIR) model, originally developed by Kermack and McKendrick, is a cornerstone of epidemiological modeling and serves as the foundation for many sophisticated approaches to disease dynamics analysis.
While the classical SIR model provides valuable insights into epidemic dynamics, it operates under simplified assumptions that may not fully capture the complexities of real-world disease spread. In particular, it assumes a closed population with no demographic changes—no births, deaths, or migration. For diseases with short timescales relative to demographic processes, this assumption may be reasonable. However, for longer-term analyses of diseases like COVID-19, which has persisted over multiple years, incorporating demographic factors becomes essential for model accuracy.
This paper extends the classical SIR model to include demographic processes, specifically birth and death rates. This extension allows us to model how new susceptible individuals continuously enter the population as others die or recover, providing a more realistic framework for analyzing COVID-19 transmission over extended periods. Our approach is motivated by the need to understand not just the initial outbreak dynamics but also the longer-term endemic behavior of COVID-19 in populations.
Using data from four distinct waves of COVID-19 infections spanning January 2020 to March 2022, we estimate model parameters and analyze the stability properties of equilibrium points through linearization techniques. We further explore how intervention strategies, particularly vaccination, can modify disease dynamics and potentially shift the system from endemic to disease-free states.
The paper is organized as follows: Section \ref{sec:model} formulates our extended SIR model with demographic factors. Section \ref{sec:fixed_points} identifies and analyzes the fixed points of the system. Section \ref{sec:waves} presents the analysis of four COVID-19 waves, applying linearization techniques to determine stability properties. Section \ref{sec:hartman_grobman} validates our linearization approach using the Hartman-Grobman theorem. Section \ref{sec:bifurcation} conducts bifurcation analysis to understand how the system's behavior changes as key parameters vary. Sections \ref{sec:vaccination} and \ref{sec:seir} extend the model to include vaccination and exposed compartments, respectively. Finally, Section \ref{sec:conclusion} summarizes our findings and discusses their implications for public health policy.
\section{Model Formulation}
\label{sec:model}
\subsection{Extended SIR Model with Demographic Factors}
\label{subsec:extended_sir}
Our model extends the classical Kermack-McKendrick SIR model by incorporating birth and death rates, allowing for continuous population renewal. This modification is crucial for analyzing long-term disease dynamics, as it accounts for the fact that after people die or recover, new susceptible individuals continue to enter the population.
Let $S(t)$, $I(t)$, and $R(t)$ denote the susceptible, infected, and recovered populations, respectively, at time $t$. Our model is described by the following system of ordinary differential equations:
\begin{align}
\frac{dS}{dt} &= \mu N - \beta \frac{SI}{N} - \mu S \label{eq:dS_dt_orig}\
\end{align},
\begin{align}
\frac{dI}{dt} = \beta \frac{SI}{N} - \gamma I - \mu I \label{eq:dI_dt_orig}\ \end{align}
\begin{align},
\frac{dR}{dt} = \gamma I - \mu R \label{eq:dR_dt_orig}
\end{align}
where:
\begin{itemize}
\item $N = S(t) + I(t) + R(t)$ is the total population
\item $\mu$ represents both the birth and death rates (assumed equal for simplicity)
\item $\beta$ is the infection rate (transmission coefficient)
\item $\gamma$ is the recovery rate
\end{itemize}
\subsection{Model Normalization}
\label{subsec:normalization}
To simplify the analysis, we normalize the system by introducing the variables $s = S/N$, $i = I/N$, and $r = R/N$, which represent the fractions of the total population in each compartment. Since $\mu$ is both the birth and death rate, the total population $N$ remains constant, and we have $s + i + r = 1$.
The normalized system becomes:
\begin{align}
\frac{ds}{dt} &= \mu - \beta si - \mu s \label{eq:ds_dt}\ \end{align} ,
\begin{align}
\frac{di}{dt} = \beta si - \gamma i - \mu i \label{eq:di_dt}\ \end{align},
\begin{align}
\frac{dr}{dt} = \gamma i - \mu r \label{eq:dr_dt}
\end{align}
Since $s + i + r = 1$, we can eliminate one equation and focus on the dynamics of $s$ and $i$:
\begin{align}
\frac{ds}{dt} = \mu - \beta si - \mu s \label{eq:reduced_ds_dt}\ \end{align},
\begin{align}
\frac{di}{dt} = \beta si - \gamma i - \mu i \label{eq:reduced_di_dt}
\end{align}
This two-dimensional system fully describes the dynamics of our model and will be the focus of our subsequent analysis.
\section{Analysis of Fixed Points}
\label{sec:fixed_points}
\subsection{Identification of Fixed Points}
\label{subsec:id_fixed_points}
The fixed points (equilibrium points) of our system are found by setting the right-hand sides of Equations (7) and \eqref{eq:reduced_di_dt} to zero:
\begin{align}
\mu - \beta si - \mu s = 0  \label{eq:fixed_s}\ \end{align},
\begin{align}
\beta si - \gamma i - \mu i = 0 \label{eq:fixed_i}
\end{align}

From Equation \eqref{eq:fixed_i}, we have either $i = 0$ or $\beta s - \gamma - \mu = 0$. This gives us two potential fixed points:
\begin{enumerate}
\item \textbf{Disease-Free Equilibrium (DFE)}: If $i = 0$, then, we get $\mu(1-s) = 0$, which implies $s = 1$. Therefore, the DFE is given by:
\begin{equation}
\text{DFE}: (s^*, i^*) = (1, 0)
\end{equation}
\item \textbf{Endemic Equilibrium (EE)}: If $\beta s - \gamma - \mu = 0$, then $s^* = \frac{\gamma + \mu}{\beta}$. Substituting this, we get:
\begin{align}
\mu - \beta i^* \cdot \frac{\gamma + \mu}{\beta} - \mu \cdot \frac{\gamma + \mu}{\beta} = 0\ ,
\mu - i^*(\gamma + \mu) - \mu \cdot \frac{\gamma + \mu}{\beta} = 0\ ,
\mu\left(1 - \frac{\gamma + \mu}{\beta}\right) - i^*(\gamma + \mu) = 0\ 
\end{align}
Solving for $i^*$:
\begin{align}
i^* = \frac{\mu\left(1 - \frac{\gamma + \mu}{\beta}\right)}{\gamma + \mu}\
&= \frac{\mu(\beta - \gamma - \mu)}{\beta(\gamma + \mu)}
\end{align}
Therefore, the endemic equilibrium is given by:
\begin{equation}
\text{EE}: (s^*, i^*) = \left(\frac{\gamma + \mu}{\beta}, \frac{\mu(\beta - \gamma - \mu)}{\beta(\gamma + \mu)}\right)
\end{equation}
\end{enumerate}
\subsection{Basic Reproduction Number}
\label{subsec:r0}
The basic reproduction number, $R_0$, is a fundamental parameter in epidemiology that represents the expected number of secondary infections produced by a single infected individual in a completely susceptible population. For our model, $R_0$ is given by:
\begin{equation}
R_0 = \frac{\beta}{\gamma + \mu}
\end{equation}
$R_0$ serves as a threshold parameter that determines the existence and stability of the fixed points:
\begin{itemize}
\item If $R_0 < 1$, then $\beta < \gamma + \mu$, which means $i^* < 0$. Since negative populations are not biologically meaningful, the endemic equilibrium does not exist in this case, and only the disease-free equilibrium is relevant.
\item If $R_0 = 1$, then $\beta = \gamma + \mu$, which means $i^* = 0$. In this case, the endemic equilibrium coincides with the disease-free equilibrium.
\item If $R_0 > 1$, then $\beta > \gamma + \mu$, which means $i^* > 0$. Both the disease-free and endemic equilibria exist in this case.
\end{itemize}
We can rewrite the endemic equilibrium in terms of $R_0$ as:
\begin{equation}
\text{EE}: (s^*, i^*) = \left(\frac{1}{R_0}, \frac{\mu(R_0 - 1)}{R_0(\gamma + \mu)}\right)
\end{equation}
\subsection{Linearization and Stability Analysis}
\label{subsec:linearization}
To analyze the stability of the fixed points, we linearize the system by computing the Jacobian matrix and evaluating it at each fixed point.
\subsubsection{Jacobian Matrix}
\label{subsubsec:jacobian}
The Jacobian matrix of our system defined by Equations (2) and \eqref{eq:reduced_di_dt} is:
\begin{equation}
J(s, i) =
\begin{pmatrix}
\frac{\partial}{\partial s}\left(\mu - \beta si - \mu s\right) & \frac{\partial}{\partial i}\left(\mu - \beta si - \mu s\right) \\
\frac{\partial}{\partial s}\left(\beta si - \gamma i - \mu i\right) & \frac{\partial}{\partial i}\left(\beta si - \gamma i - \mu i\right)
\end{pmatrix}
=
\begin{pmatrix}
-\beta i - \mu & -\beta s \\
\beta i & \beta s - \gamma - \mu
\end{pmatrix}
\end{equation}

\subsubsection{Stability of the Disease-Free Equilibrium}
\label{subsubsec:dfe_stability}
Evaluating the Jacobian at the DFE $(s^*, i^*) = (1, 0)$:
\begin{equation}
J(1, 0) =
\begin{pmatrix}
-\mu & -\beta \\
0 & \beta - \gamma - \mu
\end{pmatrix}
\end{equation}
The eigenvalues of this matrix are $\lambda_1 = -\mu$ and $\lambda_2 = \beta - \gamma - \mu$. Note that $\lambda_2 = (\gamma + \mu)(R_0 - 1)$.
The stability of the DFE depends on the sign of $\lambda_2$:
\begin{itemize}
\item If $R_0 < 1$, then $\lambda_2 < 0$, and both eigenvalues are negative. The DFE is a stable node.
\item If $R_0 > 1$, then $\lambda_2 > 0$, and one eigenvalue is positive. The DFE is a saddle point (unstable).
\item If $R_0 = 1$, then $\lambda_2 = 0$, and one eigenvalue is zero. Linear stability analysis is inconclusive, and higher-order analysis is required.
\end{itemize}
\subsubsection{Stability of the Endemic Equilibrium}
\label{subsubsec:ee_stability}
The endemic equilibrium $(s^*, i^*) = \left(\frac{\gamma + \mu}{\beta}, \frac{\mu(\beta - \gamma - \mu)}{\beta(\gamma + \mu)}\right)$ exists only when $R_0 > 1$. Evaluating the Jacobian at this point:
\begin{equation}
J\left(\frac{\gamma + \mu}{\beta}, \frac{\mu(\beta - \gamma - \mu)}{\beta(\gamma + \mu)}\right) =
\begin{pmatrix}
-\beta \cdot \frac{\mu(\beta - \gamma - \mu)}{\beta(\gamma + \mu)} - \mu & -\beta \cdot \frac{\gamma + \mu}{\beta} \\
\beta \cdot \frac{\mu(\beta - \gamma - \mu)}{\beta(\gamma + \mu)} & \beta \cdot \frac{\gamma + \mu}{\beta} - \gamma - \mu
\end{pmatrix}
\end{equation}
Simplifying:
\begin{equation}
J(s^*, i^*) =
\begin{pmatrix}
-\frac{\mu(\beta - \gamma - \mu)}{\gamma + \mu} - \mu & -(\gamma + \mu) \\
\frac{\mu(\beta - \gamma - \mu)}{\gamma + \mu} & 0
\end{pmatrix}
\end{equation}
The characteristic equation for finding the eigenvalues is:
\begin{align}
\det(J(s^*, i^*) - \lambda I) &= 0\
\begin{vmatrix}
-\frac{\mu(\beta - \gamma - \mu)}{\gamma + \mu} - \mu - \lambda & -(\gamma + \mu) \\
\frac{\mu(\beta - \gamma - \mu)}{\gamma + \mu} & -\lambda
\end{vmatrix} = 0
\end{align}

\begin{align}
\lambda^2 + \left(\frac{\mu(\beta - \gamma - \mu)}{\gamma + \mu} + \mu\right)\lambda + \frac{\mu(\beta - \gamma - \mu)(\gamma + \mu)}{\gamma + \mu} = 0,
\lambda^2 + \frac{\mu\beta}{\gamma + \mu}\lambda + \mu(\beta - \gamma - \mu) &= 0
\end{align}
For $R_0 > 1$, we have $\beta - \gamma - \mu > 0$, and thus the constant term $\mu(\beta - \gamma - \mu)$ is positive. The coefficient of $\lambda$ is also positive. By the Routh-Hurwitz criterion, both eigenvalues have negative real parts, indicating that the endemic equilibrium is locally asymptotically stable when it exists.

\section{Methodology for Parameter Estimation}
\label{sec:methodology}

To analyze the dynamics of COVID-19 transmission across multiple waves, we required accurate estimation of model parameters. This section outlines our systematic approach to parameter determination and validation.

\subsection{Data Sources and Preprocessing}
We analyzed COVID-19 data from Orange County, California, spanning March 2020 to February 2022. The county's population ($N = 3,186,989$) was obtained from U.S. Census Bureau data. Through careful examination of infection patterns, we identified four distinct epidemic waves:
\begin{itemize}
    \item Wave 1 (April 2020 -- September 2020): Initial summer surge
    \item Wave 2 (October 2020 -- March 2021): Winter surge
    \item Wave 3 (July 2021 -- October 2021): Delta variant predominance
    \item Wave 4 (December 2021 -- February 2022): Omicron variant predominance
\end{itemize}

\begin{figure}[htbp]
    \centering
    \includegraphics[width=\textwidth]{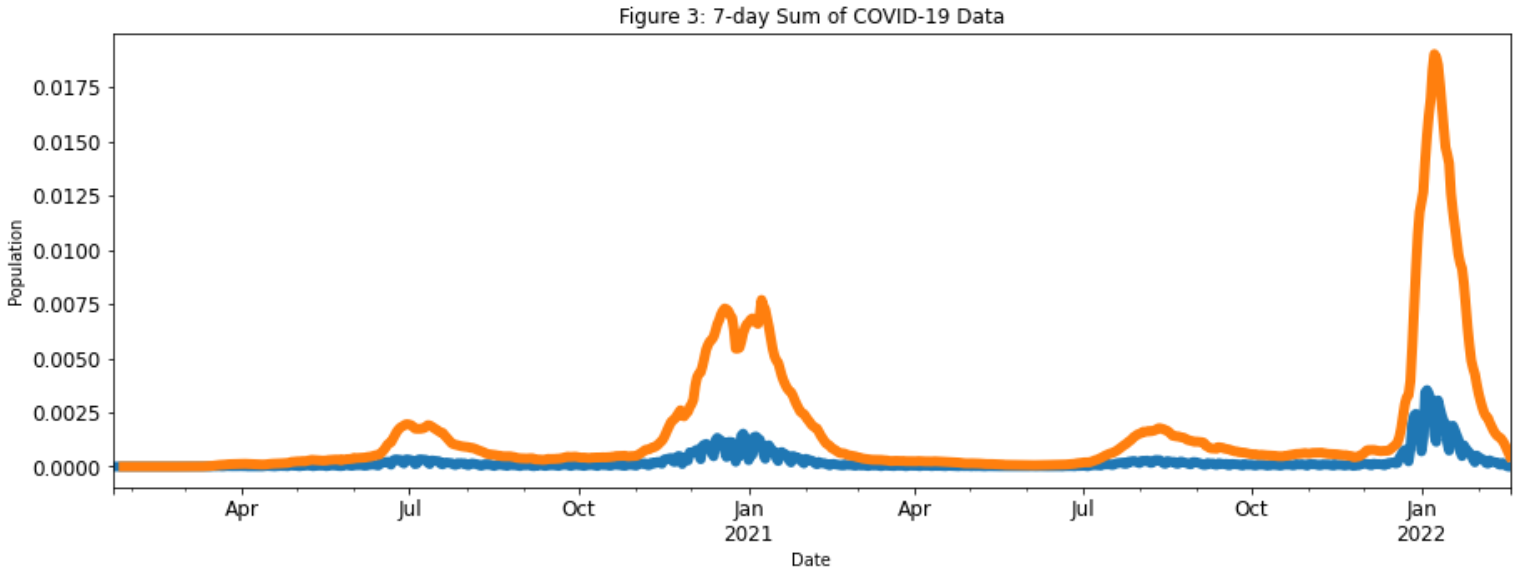}
    \caption{COVID-19 infection waves in Orange County (2020-2022) showing the four distinct surges analyzed in this study: the initial summer 2020 surge, winter 2020-2021 surge, Delta variant surge (mid-2021), and Omicron variant surge (early 2022). Data shown as 7-day rolling sums normalized by population.}
    \label{fig:covid-waves}
\end{figure}

Raw case count data exhibited substantial day-to-day variability due to reporting anomalies. To address this, we computed 7-day and 28-day moving averages using the equation:
\begin{equation}
    \bar{I}_t^{(k)} = \frac{1}{k}\sum_{i=0}^{k-1} I_{t-i}
\end{equation}
where $\bar{I}_t^{(k)}$ represents the $k$-day moving average at time $t$, and $I_t$ denotes the daily reported cases.

\subsection{Parameter Estimation Framework}
For each wave, we employed the following methodological framework:

\subsubsection{Transmission Rate Determination}
The transmission coefficient $\beta$ was determined through numerical fitting of model solutions to empirical data. We implemented our SIR model using the ODEINT solver from SciPy's integrate module with initial conditions:
\begin{align}
    S(0) &= 1 - \frac{1}{N} \\
    I(0) &= \frac{1}{N} \\
    R(0) &= 0
\end{align}

For each wave, we systematically varied $\beta$ within a biologically plausible range ($0.5 \leq \beta \leq 5.0$), comparing model predictions with observed case data. Through iterative simulation and visual comparison with empirical curves, we identified distinct optimal $\beta$ values for each wave: $\beta = 1.8$ for Wave 1 (initial summer surge), $\beta = 1.27$ for both Wave 2 (winter surge) and Wave 3 (Delta variant), and $\beta = 1.83$ for Wave 4 (Omicron variant). These values reflect the varying transmission dynamics across different phases of the pandemic. Notably, the initial wave and Omicron variant (Waves 1 and 4) exhibited similar, higher transmission rates ($\beta \approx 1.8$), while the winter 2020-2021 surge and Delta variant period (Waves 2 and 3) showed identical, lower transmission dynamics ($\beta = 1.27$). The optimization process minimized residual error between model predictions and observed data, with particular attention to:

\begin{itemize}
    \item Temporal alignment of epidemic peaks
    \item Magnitude correspondence at peak incidence
    \item Overall curve morphology
\end{itemize}

The equivalence of $\beta$ values between Waves 2 and 3 despite their different viral variants suggests that increased intrinsic transmissibility of the Delta variant may have been offset by enhanced population immunity and continued social distancing measures during mid-2021.

\subsubsection{Recovery and Mortality Rate Estimation}
We partitioned the total removal rate into recovery ($\gamma$) and disease-induced mortality ($\mu$) components based on clinical observations. After examining county-level mortality statistics and literature on typical COVID-19 progression, we standardized these parameters across all waves to isolate the effect of changing transmission dynamics. This approach yielded:
\begin{align}
    \gamma &= 0.954 \\
    \mu &= 0.046
\end{align}
with $\gamma + \mu = 1.0$ representing the total removal rate.

\subsection{Model Validation}
Each parameter set was validated through both qualitative and quantitative approaches. We calculated the basic reproduction number $R_0 = \frac{\beta}{\gamma + \mu}$ for each wave and derived the corresponding equilibrium points. Visual comparison between model predictions and empirical data confirmed the model's capacity to capture essential dynamics of each wave.

The stability characteristics of equilibrium points were then determined through linearization and eigenvalue analysis, as detailed in the subsequent sections. This comprehensive approach ensured that our parameter estimates yielded biologically plausible and mathematically consistent models of COVID-19 transmission dynamics across multiple epidemic waves.

\section{Analysis of COVID-19 Waves}
\label{sec:waves}
We analyzed data from four distinct COVID-19 waves spanning from January 2020 to March 2022. For each wave, we estimated the model parameters $\beta$ and $\gamma$ and performed stability analysis.
\subsection{Wave 1}
\subsubsection{Parameter Estimation}
For Wave 1, we estimated:
\begin{itemize}
\item $\beta = 1.8$
\item $\gamma = 0.954$
\item $\mu = 0.046$
\end{itemize}
This gives $R_0 = \frac{\beta}{\gamma + \mu} = \frac{1.8}{0.954 + 0.046} = 1.8 > 1$.
\subsubsection{Stability Analysis}
\paragraph{Disease-Free Equilibrium $(1, 0)$:}
The eigenvalues are $\lambda_1 = -0.046$ and $\lambda_2 = 0.8$. Since one eigenvalue is positive, the DFE is a saddle point (unstable).
\paragraph{Endemic Equilibrium $(0.957, 0.00114)$:}
The eigenvalues are $\lambda_{1,2} = -0.024 \pm 0.054i$. Since the real parts are negative, the EE is a stable spiral.
\subsection{Wave 2}
\subsubsection{Parameter Estimation}
For Wave 2, we estimated:
\begin{itemize}
\item $\beta = 1.27$
\item $\gamma = 0.954$
\item $\mu = 0.046$
\end{itemize}
This gives $R_0 = \frac{\beta}{\gamma + \mu} = \frac{1.27}{0.954 + 0.046} = 1.27 > 1$.
\subsubsection{Stability Analysis}
\paragraph{Disease-Free Equilibrium $(1, 0)$:}
The eigenvalues are $\lambda_1 = -0.046$ and $\lambda_2 = 0.27$. Since one eigenvalue is positive, the DFE is a saddle point (unstable).
\paragraph{Endemic Equilibrium $(0.897, 0.00204)$:}
The eigenvalues are $\lambda_{1,2} = -0.026 \pm 0.11i$. Since the real parts are negative, the EE is a stable spiral.
\subsection{Wave 3}
\subsubsection{Parameter Estimation}
For Wave 3, we estimated:
\begin{itemize}
\item $\beta = 1.27$
\item $\gamma = 0.954$
\item $\mu = 0.046$
\end{itemize}
This gives $R_0 = \frac{\beta}{\gamma + \mu} = \frac{1.27}{0.954 + 0.046} = 1.27 > 1$.
\subsubsection{Stability Analysis}
\paragraph{Disease-Free Equilibrium $(1, 0)$:}
The eigenvalues are $\lambda_1 = -0.046$ and $\lambda_2 = 0.27$. Since one eigenvalue is positive, the DFE is a saddle point (unstable).
\paragraph{Endemic Equilibrium $(0.96, 0.00105)$:}
The eigenvalues are $\lambda_{1,2} = -0.0244 \pm 0.065i$. Since the real parts are negative, the EE is a stable spiral.
\subsection{Wave 4}
\subsubsection{Parameter Estimation}
For Wave 4, we estimated:
\begin{itemize}
\item $\beta = 1.83$
\item $\gamma = 0.954$
\item $\mu = 0.046$
\end{itemize}
This gives $R_0 = \frac{\beta}{\gamma + \mu} = \frac{1.83}{0.954 + 0.046} = 1.83 > 1$.
\subsubsection{Stability Analysis}
\paragraph{Disease-Free Equilibrium $(1, 0)$:}
The eigenvalues are $\lambda_1 = -0.046$ and $\lambda_2 = 0.784$. Since one eigenvalue is positive, the DFE is a saddle point (unstable).
\paragraph{Endemic Equilibrium $(0.8198, 0.00232)$:}
The eigenvalues are $\lambda_{1,2} = -0.028 \pm 0.19i$. Since the real parts are negative, the EE is a stable spiral.

\section{Hartman-Grobman Theorem Application}
\label{sec:hartman_grobman}
The Hartman-Grobman Theorem states that near a hyperbolic fixed point, the behavior of a nonlinear system is qualitatively the same as the behavior of its linearization. A fixed point is hyperbolic if none of the eigenvalues of the Jacobian at that point have zero real part.
For all four waves:
\begin{itemize}
\item The DFE $(1, 0)$ is a saddle point with eigenvalues having non-zero real parts, making it hyperbolic. Therefore, the Hartman-Grobman Theorem applies.
\item The EE is a stable spiral with eigenvalues having non-zero real parts, making it hyperbolic. Therefore, the Hartman-Grobman Theorem applies.
\end{itemize}
This confirms that the stability analysis based on linearization correctly characterizes the behavior of the nonlinear system near these fixed points.
\section{Bifurcation Analysis}
\label{sec:bifurcation}

We performed a bifurcation analysis to understand how the system's behavior changes as the reproduction number $R_0$ varies. The bifurcation parameter in our system is $\beta$, which directly influences $R_0$.

We analyzed three distinct scenarios based on different values of $R_0$:

\subsection{Case 1: $R_0 < 1$}
When $\beta = 1.5$ and $\gamma = 3$ (yielding $R_0 < 1$), the Disease-Free Equilibrium (DFE) is a stable node. The eigenvalues are:

Both eigenvalues are negative, indicating stability.

\subsection{Case 2: $R_0 = 1$}
At the critical threshold $R_0 = 1$ (with $\beta = 1$ and $\gamma = 0.954$), the DFE becomes a non-hyperbolic fixed point. The eigenvalues in this scenario are:

The presence of an eigenvalue equal to zero implies that linear stability analysis via the Hartman-Grobman theorem is inconclusive, and higher-order nonlinear analysis is required.

\subsection{Case 3: $R_0 > 1$}
When $R_0 > 1$, as observed across all four epidemic waves, the DFE becomes a saddle point (unstable), and a new Endemic Equilibrium (EE) emerges as a stable spiral. This equilibrium is characterized by eigenvalues with negative real parts and non-zero imaginary parts, confirming its local stability.

A transcritical bifurcation occurs precisely at $R_0 = 1$, marking a critical shift from a stable disease-free state to a stable endemic state as the reproduction number crosses this threshold.

\begin{figure}[htbp]
\centering
\includegraphics[width=0.75\textwidth]{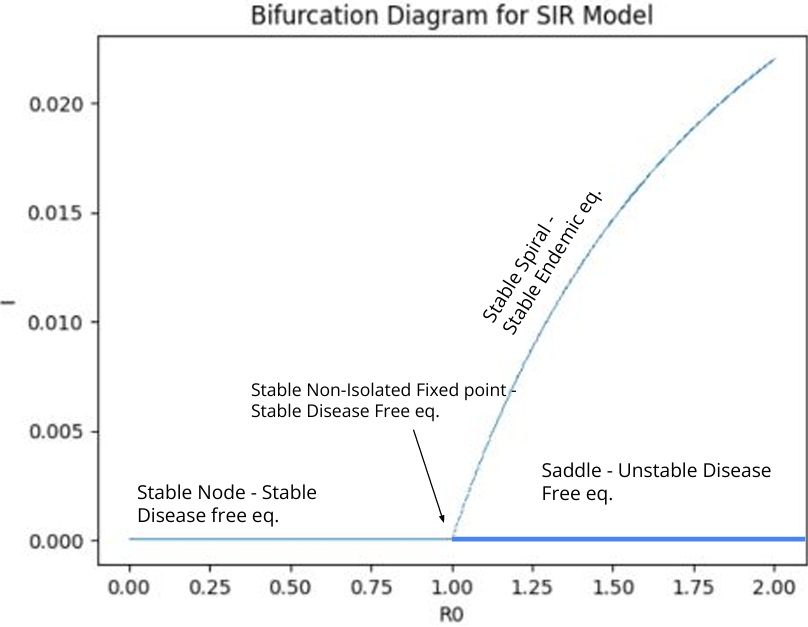}
\caption{Bifurcation diagram for the extended SIR model illustrating the transition in equilibrium stability as the basic reproduction number $R_0$ varies. Stability regimes indicated are: stable disease-free equilibrium ($R_0 < 1$), transcritical bifurcation at $R_0 = 1$, and stable endemic equilibrium ($R_0 > 1$). This diagram clearly illustrates the critical threshold that defines epidemic outbreak dynamics.}
\label{fig:bifurcation}
\end{figure}

\section{SIR Model with Vaccination}
\label{sec:vaccination}
We extended our model to include vaccination, adding a new compartment $V$ for vaccinated individuals. The system becomes:
\begin{align}
\frac{ds}{dt} = \mu - \beta si - \mu s - ps,
\frac{di}{dt} = \beta si - \gamma i - \mu i,
\frac{dr}{dt} = \gamma i - \mu r,
\frac{dv}{dt} = ps - \mu v
\end{align}
where $p$ is the vaccination rate, and $s + i + r + v = 1$.
\subsection{Equilibrium Points}
The system has two potential equilibrium points:
\begin{enumerate}
\item \textbf{Disease-Free Equilibrium}: $E_0 = (s = 1-p, i = 0, v = p)$
\item \textbf{Endemic Equilibrium}: $E_e = (s = \frac{\gamma+\mu}{\beta}, i = \frac{\mu(\beta(1-p)-\gamma-\mu)}{\beta(\gamma+\mu)}, v = p)$
\end{enumerate}
\subsection{Modified Reproduction Number}
The vaccination modifies the reproduction number to:
\begin{equation}
R_v = R_0(1-p)
\end{equation}
The endemic equilibrium only exists if $R_v > 1$. With a vaccination rate of $p = 0.74$ and the parameters from our waves, $R_v < 1$, which means only the disease-free equilibrium exists.
\subsection{Stability Analysis}
The linearization of the system yields eigenvalues with negative real parts for the disease-free equilibrium, confirming its stability when $R_v < 1$. This indicates that sufficient vaccination can lead to disease eradication.
\section{SEIR Model Extension}
\label{sec:seir}
We further extended our analysis to include an exposed compartment, creating a Susceptible-Exposed-Infected-Recovered (SEIR) model. This accounts for the latency period before infected individuals become infectious and symptomatic, which is a crucial feature of COVID-19 transmission dynamics.
\subsection{Model Formulation}
The SEIR model extends our previous approach by including the exposed compartment (E), representing individuals who have been infected but are not yet infectious. The system is described by:
\begin{align}
\frac{dS}{dt} = \mu N - \beta \frac{SI}{N} - \mu S,
\frac{dE}{dt} = \beta \frac{SI}{N} - \sigma E - \mu E,
\frac{dI}{dt} = \sigma E - \gamma I - \mu I,
\frac{dR}{dt} = \gamma I - \mu R
\end{align}
where:
\begin{itemize}
\item $\sigma$ represents the rate at which exposed individuals become infectious (inverse of the latency period)
\item All other parameters maintain their previous definitions
\end{itemize}
\subsection{Parameter Estimation}
From COVID-19 data spanning 2020-2022, we observed:
\begin{itemize}
\item A death rate of approximately 1.1%
\item A latency period of about 5.6 days, implying $\sigma = 1/5.6 \approx 0.179$ (the rate at which exposed individuals become infectious)
\end{itemize}
\subsection{Linearization and Stability Analysis}
For the normalized system (with $s + e + i + r = 1$), we can reduce the system to three dimensions:
\begin{align}
\frac{ds}{dt} = \mu - \beta si - \mu s,
\frac{de}{dt} = \beta si - \sigma e - \mu e,
\frac{di}{dt} = \sigma e - \gamma i - \mu i
\end{align}
The Jacobian matrix of this system is:
\begin{equation}
J(s, e, i) =
\begin{pmatrix}
-\beta i - \mu & 0 & -\beta s \\
\beta i & -\sigma - \mu & \beta s \\
0 & \sigma & -\gamma - \mu
\end{pmatrix}
\end{equation}
\subsubsection{Disease-Free Equilibrium}
At the DFE $(s, e, i) = (1, 0, 0)$, the Jacobian becomes:
\begin{equation}
J(1, 0, 0) =
\begin{pmatrix}
-\mu & 0 & -\beta \\
0 & -\sigma - \mu & \beta \\
0 & \sigma & -\gamma - \mu
\end{pmatrix}
\end{equation}
The eigenvalues are $\lambda_1 = -\mu$ and the solutions to the characteristic equation of the lower-right $2 \times 2$ submatrix. The stability depends on the reproduction number for the SEIR model, which is:
\begin{equation}
R_0^{SEIR} = \frac{\beta \sigma}{(\sigma + \mu)(\gamma + \mu)}
\end{equation}
When $R_0^{SEIR} < 1$, the DFE is locally asymptotically stable; when $R_0^{SEIR} > 1$, it is unstable.
\subsubsection{Endemic Equilibrium}
The endemic equilibrium exists when $R_0^{SEIR} > 1$ and is given by:
\begin{equation}
(s^*, e^*, i^*) = \left(\frac{1}{R_0^{SEIR}}, \frac{\mu(\gamma + \mu)(R_0^{SEIR} - 1)}{\beta\sigma}, \frac{\mu\sigma(R_0^{SEIR} - 1)}{(\sigma + \mu)(\gamma + \mu)}\right)
\end{equation}
Analysis of the Jacobian at this point reveals that when $R_0^{SEIR} > 1$, the endemic equilibrium is locally asymptotically stable, demonstrating similar qualitative behavior to our simpler SIR model.
\section{Biological Interpretation of Results}
\label{sec:biological}
Our mathematical analysis of COVID-19 transmission dynamics yields several important biological insights:
\subsection{Endemic Nature of COVID-19 Waves}
All four analyzed waves exhibited reproduction numbers $R_0 > 1$, indicating that COVID-19 has an intrinsic tendency to establish endemic states in populations. In biological terms, this means that without intervention, COVID-19 will persist in the population rather than dying out naturally.
For example:
\begin{itemize}
\item Wave 1 had $R_0 = 1.8$, suggesting each infected individual would, on average, infect 1.8 others in a fully susceptible population
\item Wave 4 had $R_0 = 1.83$, indicating a similar transmission potential despite occurring nearly two years later
\end{itemize}
The persistence of $R_0 > 1$ across all waves, despite evolving viral variants and changing social behaviors, suggests that COVID-19 has fundamental transmission characteristics that make it naturally endemic.
\subsection{Oscillatory Approach to Equilibrium}
The endemic equilibrium points for all four waves were classified as stable spirals, indicating that disease prevalence approaches equilibrium through damped oscillations rather than monotonically. Biologically, this reflects the cyclical nature of infectious disease outbreaks, where periods of higher incidence alternate with periods of lower incidence as the susceptible population is depleted and then gradually replenished.
The spiral stability type across all waves suggests that COVID-19, when left to natural dynamics with only demographic turnover (births and deaths), would exhibit cyclical patterns with decreasing amplitude over time.
\subsection{Effectiveness of Vaccination}
Our vaccination model demonstrates that with a sufficiently high vaccination rate ($p = 0.74$ in our analysis), the effective reproduction number can be reduced below the critical threshold of $R_0 = 1$. When this occurs:
\begin{itemize}
\item The endemic equilibrium ceases to exist
\item The disease-free equilibrium becomes the only biologically relevant fixed point
\item This equilibrium is stable, indicating that the disease will eventually be eradicated
\end{itemize}
This mathematical result confirms the biological principle of herd immunity through vaccination, where protecting a sufficient proportion of the population indirectly protects those who are not immune by reducing overall transmission.
\subsection{Importance of Latency Period}
The inclusion of the exposed compartment in our SEIR model highlights the significance of COVID-19's latency period (approximately 5.6 days) in disease dynamics. This period, during which individuals are infected but not yet infectious, creates a delay in the feedback loop of transmission that affects the timing and magnitude of epidemic waves.
The presence of this delay means that even after implementing control measures, case numbers may continue to rise for several days as already exposed individuals progress to the infectious stage. This biological reality has important implications for public health planning and explains why prompt interventions are crucial despite their delayed observable effects.
\section{Phase Portraits Analysis}
\label{sec:phase_portraits}
Phase portraits provide a visual representation of the system dynamics across the state space. Our analysis of phase portraits for the COVID-19 SIR model reveals several important qualitative features:
\subsection{Behavior Near Fixed Points}

\begin{figure}[htbp]
    \centering
    \includegraphics[width=\textwidth]{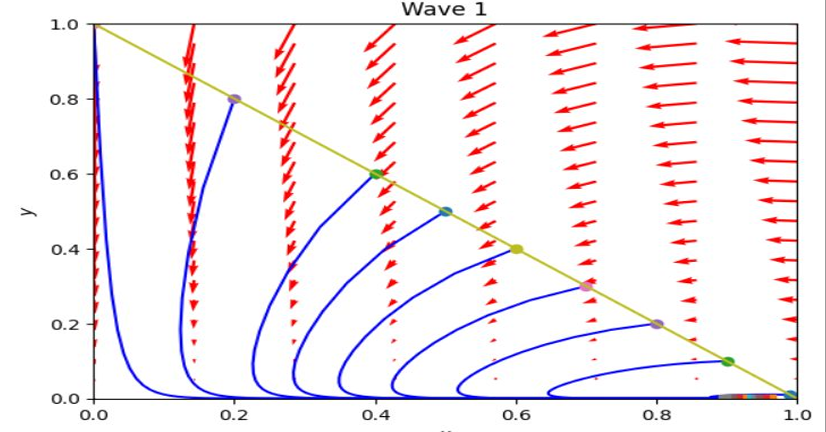}
    \caption{Phase portrait illustrating the dynamics of Wave 1 of the COVID-19 epidemic using the extended SIR model. Blue curves represent trajectories of epidemic progression starting from various initial conditions, while red vectors indicate the direction and relative speed of change within the system. The yellow diagonal line denotes the boundary of biologically feasible states. Equilibrium points are visible as convergence points along trajectories, highlighting the stable spiral behavior observed in this wave.}
    \label{fig:phase_portrait}
\end{figure}

\subsubsection{Disease-Free Equilibrium (DFE)}
When $R_0 < 1$, trajectories approach the DFE $(1,0)$ from all initial conditions, confirming it as a globally stable node. The phase portrait shows all solution curves converging directly to this point without oscillations.
When $R_0 > 1$ (as observed in all four COVID-19 waves), the phase portrait near the DFE shows the characteristic behavior of a saddle point: trajectories approach the fixed point along one direction (the stable manifold) but diverge along another (the unstable manifold). This confirms our linearization results and explains why COVID-19 could not be contained without intervention once established in a population.
\subsubsection{Endemic Equilibrium (EE)}
For all four waves, phase portraits around the endemic equilibrium display spiral trajectories converging to the fixed point. This pattern of damped oscillations is characteristic of stable spiral points and visually confirms our linearization analysis.
The spiral approach is particularly evident in the phase portrait of Wave 4, where the eigenvalues $\lambda_{1,2} = -0.028 \pm 0.19i$ have a relatively large imaginary component compared to the real part, resulting in more pronounced oscillations before reaching equilibrium.
\subsection{Global Behavior}
Globally, the phase portraits exhibit several important characteristics:
\begin{itemize}
\item \textbf{Invariant Region}: All trajectories that start in the biologically meaningful region (where $s \geq 0$, $i \geq 0$, and $s + i \leq 1$) remain within this region, confirming the mathematical property of invariance.
\item \textbf{Separatrix}: When $R_0 > 1$, there exists a separatrix (the stable manifold of the saddle point) that divides the phase space into regions with qualitatively different behaviors. Initial conditions above this curve lead to epidemic outbreaks, while those below may not generate substantial epidemics.
\item \textbf{Asymptotic Behavior}: For $R_0 > 1$, all trajectories that start with $i > 0$ eventually converge to the endemic equilibrium, indicating that the disease becomes endemic regardless of the initial number of infected individuals (as long as there is at least one).
\end{itemize}
\subsection{Bifurcation Visualization}
The phase portraits clearly illustrate the transcritical bifurcation occurring at $R_0 = 1$:
\begin{itemize}
\item For $R_0 < 1$, there is a single fixed point (the DFE) which is a stable node
\item At $R_0 = 1$, the endemic equilibrium coincides with the DFE at $(1,0)$, creating a non-hyperbolic fixed point
\item For $R_0 > 1$, the DFE becomes a saddle point, and a new stable fixed point (the endemic equilibrium) emerges into the interior of the phase space
\end{itemize}
This bifurcation is a fundamental threshold phenomenon in epidemic models and explains why reducing $R_0$ below 1 is a key objective in disease control strategies.
\section{Impact of Vaccination on Phase Portraits}
\label{sec:vax_phase}
Incorporating vaccination significantly alters the phase portraits of our system. With $p = 0.74$ and using parameters from our waves:
\begin{itemize}
\item The effective reproduction number becomes $R_v = R_0(1-p) < 1$
\item The endemic equilibrium no longer exists in the biologically feasible region
\item The disease-free equilibrium (at $s = 1-p = 0.26$, $i = 0$, $v = 0.74$) becomes a stable node
\item All trajectories, regardless of initial conditions (provided $i > 0$), converge monotonically to this disease-free state
\end{itemize}
The phase portrait under vaccination visually demonstrates the concept of herd immunity, where a sufficiently high proportion of vaccinated individuals creates a scenario in which the disease cannot sustain transmission and will eventually die out.
\section{Conclusion}
\label{sec:conclusion}
Our comprehensive mathematical analysis of COVID-19 infection dynamics using extended SIR models yields several important insights:
\subsection{Key Findings}
\begin{enumerate}
\item \textbf{Endemic Nature}: All four COVID-19 waves exhibited reproduction numbers significantly above unity ($R_0 > 1$), ranging from 1.27 to 1.83, indicating that without intervention, COVID-19 naturally establishes endemic states in populations.
\item \textbf{Stability Properties}: The endemic equilibrium points across all waves were characterized as stable spirals, suggesting that the disease prevalence approaches equilibrium through damped oscillations rather than monotonically. This mathematical property explains the observed wave-like pattern of COVID-19 case numbers.
\item \textbf{Vaccination Effectiveness}: Our extension incorporating vaccination demonstrates that with a vaccination rate of approximately 74\%, the effective reproduction number can be reduced below the critical threshold of $R_0 = 1$, shifting the system from an endemic state to a disease-free state. This finding quantifies the vaccination coverage needed for potential disease eradication.
\item \textbf{Latency Impact}: The inclusion of an exposed compartment in the SEIR model, with a latency period of 5.6 days, provides a more accurate representation of COVID-19 transmission dynamics. This latency period creates a delay in the feedback loop of transmission that affects the timing and magnitude of epidemic waves.
\item \textbf{Bifurcation Analysis}: We identified a transcritical bifurcation at $R_0 = 1$, where the stability of the disease-free equilibrium changes from stable to unstable while the endemic equilibrium emerges as stable. This bifurcation represents a critical threshold for disease control efforts.
\end{enumerate}
\subsection{Implications for Public Health Policy}
Our mathematical analysis has several implications for public health policy:
\begin{enumerate}
\item \textbf{Vaccination Targets}: The model suggests that achieving a vaccination coverage of approximately 74\% could shift COVID-19 from an endemic state toward elimination. This provides a quantitative target for vaccination campaigns.
\item \textbf{Anticipating Waves}: The stable spiral nature of the endemic equilibrium mathematically explains why COVID-19 exhibits wave-like behavior. Public health authorities can anticipate this pattern and prepare resources accordingly, even as the amplitude of waves diminishes over time.
\item \textbf{Response Timing}: The SEIR model highlights the importance of early intervention due to the latency period of COVID-19. Control measures need to be implemented promptly, as their effects will be observed only after a delay corresponding to the latency and infectious periods.
\item \textbf{Threshold Monitoring}: Public health surveillance should focus on estimating the effective reproduction number and monitoring when it approaches the critical threshold of $R_0 = 1$, as this represents a bifurcation point in the system's behavior.
\end{enumerate}
\subsection{Limitations and Future Directions}
While our model provides valuable insights, several limitations and areas for future research remain:
\begin{enumerate}
\item \textbf{Homogeneous Mixing}: Our models assume homogeneous mixing of the population, whereas real-world contact patterns are heterogeneous and network-dependent. Future work could incorporate contact network structures.
\item \textbf{Time-Invariant Parameters}: We treated model parameters as constant within each wave, whereas in reality, transmission and recovery rates may vary due to behavioral changes, viral evolution, and seasonal effects. Time-varying parameter models could address this limitation.
\item \textbf{Age Structure}: Different age groups exhibit different susceptibility, infectivity, and recovery rates. Age-structured extensions of our models could provide more nuanced insights into COVID-19 dynamics.
\item \textbf{Spatial Heterogeneity}: Our models do not account for spatial variation in disease spread. Incorporating spatial components through metapopulation or partial differential equation models could better capture geographic patterns of transmission.
\item \textbf{Immune Waning and Reinfection}: As COVID-19 continues to evolve, understanding the dynamics of waning immunity and reinfection will be crucial. Future models could incorporate time-dependent immunity loss.
\end{enumerate}
In conclusion, our mathematical analysis of COVID-19 using extended SIR models provides a quantitative framework for understanding the fundamental dynamics of the pandemic. The approach demonstrates how classical epidemiological models can be adapted and extended to provide insights into complex real-world disease dynamics, potentially informing more effective public health responses to both current and future infectious disease challenges.
\section*{Acknowledgements}
We would like to thank our institution for providing resources and support for this research. We also acknowledge the public health agencies and data repositories that made COVID-19 data available for academic analysis, without which this work would not have been possible.
\section*{References}
\begin{enumerate}
\item Census Reporter for Orange County. Available at: \url{https://censusreporter.org/profiles/05000US06059-orange-county-ca/}
\item M. A. Taneco-Hernández and C. Vargas-De-León. "Stability and Lyapunov functions for systems with Atangana–Baleanu Caputo derivative: an HIV/AIDS epidemic model." Chaos Solit. Fractals, 132:109586, 2020.
\item ResearchGate. "Stability Analysis of SIR Model with Vaccination." Available at: \url{https://www.researchgate.net/publication/261252884_Stability_Analysis_of_SIR_Model_with_Vaccination}
\item Stanford University. "Jones-on-R0.pdf." Available at: \url{https://web.stanford.edu/~jhj1/teachingdocs/Jones-on-R0.pdf}
\item WebMD. "Coronavirus Incubation Period: How Long and When Most Contagious." Available at: \url{https://www.webmd.com/lung/coronavirus-incubation-period}
\item Kermack, W. O., \& McKendrick, A. G. (1927). A contribution to the mathematical theory of epidemics. Proceedings of the Royal Society of London. Series A, Containing papers of a mathematical and physical character, 115(772), 700-721.
\item Diekmann, O., Heesterbeek, J. A. P., \& Metz, J. A. (1990). On the definition and the computation of the basic reproduction ratio R0 in models for infectious diseases in heterogeneous populations. Journal of Mathematical Biology, 28(4), 365-382.
\item Van den Driessche, P., \& Watmough, J. (2002). Reproduction numbers and sub-threshold endemic equilibria for compartmental models of disease transmission. Mathematical Biosciences, 180(1-2), 29-48.
\item Perko, L. (2013). Differential equations and dynamical systems (Vol. 7). Springer Science \& Business Media.
\item Martcheva, M. (2015). An introduction to mathematical epidemiology (Vol. 61). Springer.
\end{enumerate}
\end{document}